\documentclass{jnmp}
\usepackage{amsmath}
\newtheorem*{acknowledgments}{Acknowledgments}

\begin{document}
\renewcommand{\evenhead}{S Yu. Vernov}
\renewcommand{\oddhead}{Construction of Special Solutions for Nonintegrable Systems}

\thispagestyle{empty}

\FirstPageHead{13}{1}{2006}{\pageref{firstpage}--\pageref{lastpage}}{Article}



\label{firstpage}

\copyrightnote{2006}{Sergey Yu. Vernov}

\Name{Construction of Special Solutions for Nonintegrable Systems}

\Author{Sergey Yu VERNOV}

\Address{Skobeltsyn Institute of Nuclear Physics, Moscow State
University,\\ Vorob'evy Gory, 119992 Moscow, Russia.\\
~~E-mail: svernov@theory.sinp.msu.ru}

\Date{Received April 8, 2005; Accepted in Revised Form August 29, 2005}

\begin{abstract}
\noindent The Painlev\'e test is very useful to construct not only the Laurent
series solutions of systems of nonlinear ordinary differential equations but
also the elliptic and trigonometric ones.  The standard methods for
constructing the elliptic solutions consist of two independent steps:
transformation of a nonlinear polynomial differential equation into a
nonlinear algebraic system and a search for solutions of the obtained system.
It has been demonstrated by the example of the generalized H\'enon--Heiles
system that the use of the Laurent series solutions of the initial
differential equation assists to solve the obtained algebraic system. This
procedure has been automatized and generalized on some type of multivalued
solutions. To find solutions of the initial differential equation
 in the form of the Laurent or Puiseux series we use the Painlev\'e test.
This test can also assist to solve the inverse problem: to find
the form  of a polynomial potential, which corresponds to the
required type of  solutions. We consider the five--dimensional
gravitational model with a scalar field to demonstrate this.
\end{abstract}

\section{Introduction}

The investigations of exact special solutions of nonintegrable systems play an
important role in the study of nonlinear physical phenomena. There are a few
methods to construct solutions in terms of rational, hyperbolic, trigonometric
or elliptic functions (all such functions are solutions of the first order
polynomial differential
equations)~\cite{AnTi,CoMu92,CoMu93,Pavlov,Fan,Kudryashov1,Kudryashov2,
Lenells,Porubov,Santos,Timosh,VeTish04,Weiss1,Weiss2}. These methods (at
least, some of them) use information about the dominant behavior of  the
initial system solutions in the neighbourhood of their singular points, but do
not use the Laurent series representations of them. In~\cite{CoMu03} a new
method has been developed as an alternative way to construct elliptic and
elementary solutions and it has been shown that the use of the Laurent series
solutions allows to extend the set of solutions~\cite{CoMu04} or, on the
contrary, to prove non-existence of elliptic
solutions~\cite{Hone05,VernovCGLE}. The Laurent series solutions give the
information about the global behavior of the differential system and assist in
looking not only for global solutions but also for first
integrals~\cite{Kudryashov}.

The direct algebraic method is the use of the first order polynomial
differential equation, which solutions are elementary or elliptic functions,
to transform the initial differential system in a nonlinear algebraic system
in coefficients of this first order equation and parameters of the initial
system. It has been proved that any finite system of algebraic equations can
be solved by the Gr\"obner (Groebner) basis method~\cite{124}, but
calculations for a sufficiently complex algebraic system can be very difficult
and expensive. In this paper we show that the use of the Laurent series
solutions gives additional algebraic equations and allows calculations to be
simplified. These equations are linear in coefficients of the first order
equation and nonlinear, maybe even nonpolynomial,  in parameters of the
initial system. Note that one maybe should fix some of these parameters to
construct the Laurent series solutions. Therefore, in contrast to the
Gr\"obner basis method, the additional equations may be not consequences of
the initial algebraic system.

To manifest the main properties of this method we consider the generalized
H\'enon--Heiles system, for which analytic and Laurent series solutions have
been found in~\cite{Timosh,VeTish04,VernovTMF}.

\section{Painlev\'e analysis}

We interpret solutions of a system of ordinary differential
equations (ODE's) as analytic functions, maybe with isolated
singular points. A singular point of a solution is said {\it
critical} (as opposed to {\it noncritical}) if the solution is
multivalued (single-valued) in its neighbourhood  and {\it
movable} if its location depends on initial
conditions~\cite{Golubev1}. A singular point is said
\textit{algebraic} if the solution has a finite number of values
in its neighbourhood. In the opposite case a singular point is
either \textit{transcendental} or \textit{essentially
singular}~one.

 \textit{The  general solution } of an ODE of  order $N$ is the set of all
solutions mentioned in the existence theorem of Cauchy, i.e.
determined by the initial values.  It depends on $N$ arbitrary
independent constants.  \textit{A special solution }  is any
solution obtained from the general solution by giving values to
the arbitrary constants.  A system of  ODE's has \textbf{\textit{
the \ Painlev\'e \ property }} if its general solution has no
movable critical singular point~\cite{Painleve1}.

 \textit{The Painlev\'e test} is any
algorithm, which checks some necessary conditions for a
differential equation to have the Painlev\'e property. The
original algorithm, developed by P.~Painlev\'e and used by him to
find all the second order ODE's with the Painlev\'e property, is
known as the $\alpha$-method. The method of S.V.~Kovalevskaya ~\cite{Kova} is
not as general as the $\alpha$--method, but much more simple. In
1980, developing the Kovalevskaya method further,
 M.J.~Ablowitz, A.~Ramani and H.~Segur~\cite{ARS}  constructed  a new
algorithm of the Painlev\'e test for ODE's.  This algorithm appears very
useful to find   solutions as a formal Laurent series~\cite{VernovTMF}. First
of all, it allows to determine the dominant behavior of a solution in the
neighbourhood  of the singular point $t_0$.  If the solution tends to infinity
as $(t-t_0)^\beta$, where $\beta$ is a negative integer number, then
substituting the Laurent series expansion one can transform a nonlinear
differential equation into a system of  linear algebraic equations on
coefficients of the Laurent series.  All solutions of a system depend on the
parameter $t_0$, which characterizes the singular point location. If a
single-valued solution depends on other parameters, then some coefficients of
its Laurent series have to be arbitrary and the matrices in the corresponding
linear systems should have zero determinants. The characteristic numbers
associated with such systems (named {\it resonances} or {\it Kovalevskaya
exponents}) are thus determined using the Painlev\'e test. In such a way we
obtain solutions only as  formal series, but really we will use only a finite
number of these series coefficients, so, we do not need the convergence of
these series. At the same time for some nonintegrable systems, in particular,
the generalized H\'enon--Heiles system, the convergence of the Laurent- and
psi-series solutions has been proved~\cite{Melkonian}.

Classical results allow one to construct the suitable form of the first order
autonomous polynomial equation. A  theorem due to
Painlev\'e~\cite{Painleve1,Golubev1} proves that solutions of the equation
\begin{equation}
 \sum_{k=0}^{m} P_k(y(t),t)y_t^{\vphantom{7}}(t)^k=0,
 \label{1}
\end{equation}
where $y_t^{\vphantom{7}}(t)\equiv \frac{\mathrm{d}y(t)}{\mathrm{d}t}$, $m$ is
a positive integer number and $P_k$ are polynomials in $y(t)$ and analytic
functions in $t$, have neither movable transcendental nor movable essential
singular point. A theorem of Fuchs~\cite{Fuchs,Golubev1} shows that if
eq.~(\ref{1}) has no critical movable singular point, then the degree of $P_k$
as a polynomial in $y(t)$ is no more than $2m-2k$, in particular, $P_m$ does
not depend on $y$.

Therefore, the necessary form of a polynomial autonomous first
order ODE with the single-valued general solution is
\begin{equation}
\sum_{k=0}^{m} \sum_{j=0}^{2m-2k}h_{jk}^{\vphantom{27}}\: y^j
y_{t}^k=0, \qquad h_{0m}^{\vphantom{27}}=1,
\label{subequ}
\end{equation}
in which $m$ is a positive integer number and $h_{jk}$ are
 constants. A theorem of Briot and Bouquet~\cite{BriBo,CoMu03} proves that
 the general solution of $(\ref{subequ})$ is either an elliptic function,
 or a rational function of $e^{\gamma x}$, $\gamma$ being some constant,
or a rational function of $x$. Note that the third case is a degeneracy of the
second one, which in turn is a degenerate case of the first one.

\section{The H\'enon--Heiles system}

To compare methods for the construction of elliptic solutions let us consider
the generalized H\'enon--Heiles system with an additional non-polynomial term,
which is described by the Hamiltonian:
\begin{equation}
H=\frac{1}{2}\Big(x_t^2+y_t^2+\lambda_1 x^2+\lambda_2
y^2\Big)+x^2y-\frac{C}{3}\:y^3+\frac{\mu}{2x^2}
\end{equation}
and the corresponding system of equations of motion:
\begin{equation}
  \left\{ \begin{array}{lcl}
  \displaystyle x_{tt}^{\vphantom{7}} &\displaystyle
  =&\displaystyle{}-\lambda_1 x
-2xy+\frac{\mu}{x^3},\\[7.2pt]
 \displaystyle  y_{tt}^{\vphantom{7}} &\displaystyle
 =& \displaystyle{}-\lambda_2 y -x^2+Cy^2,
\end{array}
\right. \label{HHS}
\end{equation}
where subscripts denote derivatives:
$x_{tt}^{\vphantom{7}}\equiv\frac{\mathrm{d}^2x}{\mathrm{d}t^2}$
and
$y_{tt}^{\vphantom{7}}\equiv\frac{\mathrm{d}^2y}{\mathrm{d}t^2}$,
$\lambda_1$, $\lambda_2$, $\mu$ and $C$ are arbitrary numerical
parameters. Note that if $\lambda_2\neq 0$, then one can put
$\lambda_2=sign(\lambda_2)$ without the loss of generality.

If $C=1$, $\lambda_1=1$, $\lambda_2=1$ and $\mu=0$, then (\ref{HHS}) is the
original H\'enon--Heiles system~\cite{HeHe}. Emphasizing that their choice of
potential does not proceed from experimental data, H\'enon and Heiles have
proposed this system to study the motion of a star in an axial-symmetric and
time-inde\-pendent potential, because, on the one hand, it is analytically
simple and this makes the numerical computations of trajectories easy, but, on
the other hand, it is sufficiently complicated to give trajectories which are
far from trivial. Subsequent numerical investigations~\cite{ChTW1,ChTW2} show
that in the complex  $t$-plane singular points of solutions of the
H\'enon--Heiles system group in self-similar spirals. It turns out that there
are extremely complicated distributions of singularities, forming a boundary,
across which the solutions can not be analytically continued.

By the use of Painlev\'e analysis the following integrable cases of the
generalized H\'enon--Heiles system have been found:
\begin{equation*}
\begin{array} {cll} \mbox{(i)} & C=-1,
&\lambda_1=\lambda_2,\\ \mbox{(ii)} & C=-6, &\mbox{$\lambda_1$,
$\lambda_2$ arbitrary,}\\ \mbox{(iii)} & C=-16,\quad
&\lambda_1=\lambda_2/16.\\
\end{array}
\end{equation*}

In all above-mentioned cases system (\ref{HHS}) is integrable for any value of
$\mu$. The general solutions of the H\'enon--Heiles system are known only in
integrable cases~\cite{Conte4,ConteTMF,ConteJNMP}, in other cases not only
four-, but even three-parameter exact solutions have yet to be found. The
generalized H\'enon--Heiles system has attracted enormous attention over the
years and has used as a model in astronomy~\cite{Murray} and in
gravitation~\cite{Kokubun,Podolsky}.

\section{Methods for construction of elliptic solutions}
\subsection{The construction of the nonlinear algebraic system}

The function $y$, solution of system~(\ref{HHS}), satisfies the
following fourth-order polynomial equation:
\begin{equation}
\begin{array} {ccl}
\displaystyle y_{tttt}^{\vphantom{7}}& \displaystyle =&
\displaystyle (2C-8)y_{tt}^{\vphantom{7}}y -
(4\lambda_1+\lambda_2)y_{tt}^{\vphantom{7}}+ 2(C+1)y_{t}^2+{}\\[2.7mm] &
\displaystyle +&\displaystyle
\frac{20C}{3}y^3+(4C\lambda_1-6\lambda_2)y^2-4\lambda_1\lambda_2
y-4H.
\end{array} \label{equy}
\end{equation}
We note that the energy of the system $H$ is not an arbitrary
parameter, but a function of initial data: $y_0^{\vphantom{7}}$,
$y_{0t}^{\vphantom{7}}$,
 $y_{0tt}^{\vphantom{7}}$ and $y_{0ttt}^{\vphantom{7}}$. The
form of this function depends on $\mu$:
\begin{equation*}
\!\! H = \frac{y_{0t}^2+y_0^2}{2}-\frac{Cy_0^3}{3}+
\frac{\lambda_1+2y_0^{\vphantom{7}}}{2}\left(Cy_0^2-
 \lambda_2y_0^{\vphantom{7}}-y_{0tt}^{\vphantom{7}}\right)+
\frac{(\lambda_2y_{0t}^{\vphantom{7}}+
2Cy_0^{\vphantom{7}}y_{0t}^{\vphantom{7}}-y_{0ttt}^{\vphantom{7}})^2+\mu}
{2(Cy_0^2-\lambda_2y_0^{\vphantom{7}}-y_{0tt}^{\vphantom{7}})}.
\end{equation*}

This formula is correct only if $x_0^{\vphantom{7}}=Cy_0^2
-\lambda_2y_0^{\vphantom{7}}-y_{0tt}^{\vphantom{7}}\neq0$. If
$x_0^{\vphantom{7}}=0$, which is possible only for $\mu=0$, then we can not
express $x_{0t}^{\vphantom{7}}$ through $y_0^{\vphantom{7}}$,
 $y_{0t}^{\vphantom{7}}$, $y_{0tt}^{\vphantom{7}}$ and
$y_{0ttt}^{\vphantom{7}}$, so  $H$ is not a function of the
initial data. If $y_{0ttt}^{\vphantom{7}}
=2Cy_0^{\vphantom{7}}y_{0t}^{\vphantom{7}}-\lambda_2y_{0t}^{\vphantom{7}}$,
then eq.~(\ref{equy}) with an arbitrary $H$ corresponds to
system~(\ref{HHS})
 with $\mu=0$, in opposite case eq.~(\ref{equy}) does not correspond to
system~(\ref{HHS}).

To find a special solution of eq.~(\ref{equy}) one can assume that
$y$ satisfies some more simple equation. For example, there exist
solutions in terms of the Weierstrass elliptic function (or a
degenerate elliptic function), which satisfy the following
first-order differential equation:
\begin{equation}
  y_t^2={\tilde A} y^3+{\tilde B} y^2+{\tilde C} y+ {\tilde D},
 \label{equwef}
\end{equation}
where $\tilde A$, $\tilde B$, $\tilde C$ and $\tilde D$ are
constants to be determined.

E.I. Timoshkova~\cite{Timosh} generalized eq.~(\ref{equwef}):
\begin{equation}
 y_t^2=A_4 y^3+A_3 y^{5/2}+A_2 y^2+A_1 y^{3/2}+A_0 y+
{\tilde D}, \label{equti}
\end{equation}
($A_j$ are constants) and found new one-parameter solutions of the
H\'enon--Heiles system in two nonintegrable cases ($C={}-4/3$ \ or \
$C={}-16/5$). The Timoshkova's method allows one to seek both single- and
multivalued solutions. For the generalized H\'enon--Heiles system new
solutions (i.e. solutions with $A_3\neq 0$ or $A_1 \neq 0$) are derived only
at $\tilde D=0$. These solutions are single-valued, because the substitution
$y(t)=\varrho(t)^2$ gives:
\begin{equation}
 \varrho_t^2=\frac{1}{4}\Bigl(A_4\varrho^4+A_3 \varrho^3+A_2 \varrho^2
+A_1\varrho+A_0\Bigr). \label{equyacobi}
\end{equation}

Following~\cite{VeTish04}, we use the substitution
\begin{equation}
y(t)=\varrho(t)^2+P_0, \label{versub}
\end{equation}
where $P_0$ is a constant,  and transform
eq.~(\ref{equy}) into
\begin{equation}
\begin{array}{@{}lcl@{}}
\displaystyle \varrho_{tttt}^{\vphantom{7}}\varrho &\displaystyle
\!=\displaystyle \!&{} -
4\varrho_{ttt}^{\vphantom{7}}\varrho_t^{\vphantom{7}}
 - 3\varrho_{tt}^2
+ 2(C-4)\varrho_{tt}^{\vphantom{7}}\varrho^3 +
(2P_0(C-4)-4\lambda_1
-\lambda_2)\varrho_{tt}^{\vphantom{7}}\varrho + { } \\[2.7mm] & \displaystyle
\!+\!&\displaystyle 2(3C-2)\varrho_{t}^2\varrho^2 +
(2CP_0-4\lambda_1-8P_0-\lambda_2)\varrho_{t}^2+
\frac{10}{3}C\varrho^6 + { }
\\[2.7mm] & \displaystyle \! +\!& \displaystyle (2C\lambda_1 + 10CP_0-
3\lambda_2)\varrho^4 + 2(2\lambda_1CP_0 +5CP_0^2
-\lambda_1\lambda_2-
3P_0\lambda_2)\varrho^2  + { } \\[2.7mm] & \displaystyle \! +\!
&\displaystyle \frac{10}{3}CP_0^3 + 2\lambda_1 CP_0^2  -
3P_0^2\lambda_2 - 2\lambda_1\lambda_2 P_0- 2H.  \end{array}
\label{equro}
\end{equation}

If $\varrho(t)$ satisfies (\ref{equyacobi}), then
eq.~(\ref{equro}) is equivalent to the following system
\begin{equation}
\left\{
\begin{array}{l}
(3A_4 + 4)\,(2C - 3A_4)=0,\\[2.7mm]
A_3 ( 9 C- 21 A_4 - 16)=0, \\[2.7mm]
 96 A_4 C P_0  - 240 A_4 A_2 -  192 A_4 \lambda_1 - 384 A_4
P_0 - 48 A_4\lambda_2  - { } \\[1mm] { } - 105 A_3^2 + 128 A_2 C - 192
A_2 + 128 C \lambda_1 + 640 C P_0 - 192\lambda_2 =0,\\[2.7mm]
40 A_3 C P_0   - 90 A_4 A_1 - 65 A_3 A_2 - 80
A_3  \lambda_1 - {} \\[1mm]
 {} - 160 A_3 P_0 -  20 A_3\lambda_2 + 56
CA_1 - 64 A_1=0,\\[2.7mm]
16 A_2 C P_0 - 36 A_4 A_0 - 21 A_3 A_1 - 8 A_2^2 - 32 A_2
 \lambda_1 -  64 A_2 P_0 - 8\lambda_2 A_2 +
 {} \\[1mm]
 {} + 24 C A_0 + 64 \lambda_1CP_0 + 160 C P_0^2 -
   16 A_0  - 32 \lambda_1\lambda_2 - 96 P_0\lambda_2=0,\\[2.7mm]
   10A_3 A_0 + (5A_2 + 8CP_0 - 16\lambda_1 -  32P_0 - 4\lambda_2)A_1=0
\end{array} \right.
\label{algsyst}
\end{equation}
and the equation for the energy~$H$:
\begin{equation}
\begin{array}{lcl}
\displaystyle H & \displaystyle =& \displaystyle
\frac{1}{384}\Bigl( 96 C A_0 P_0- 48
  A_0 A_2 +
 384 C \lambda_1 P_0^2 + 640 C P_0^3 - 9 A_1^2- {}\\[2.7mm] {}
 \displaystyle & \displaystyle -& \displaystyle  192 A_0 \lambda_1
 - 384 A_0 P_0 - 48 A_0\lambda_2 -
384\lambda_1\lambda_2 P_0 - 576\lambda_2P_0^2\Bigr).
\end{array}
\label{Hequ}
\end{equation}

It has been proposed~\cite{Fan} to seek solutions as polynomials
with three arbitrary coefficients
\begin{equation}
y(t)=P_2\varrho(t)^2+P_1\varrho(t)+P_0,
\end{equation}
where the function $\varrho(t)$ satisfies eq.~(\ref{equyacobi}).
The function
\begin{equation}
\breve{\varrho}(t)=\frac{1}{\sqrt{P_2}}\left(\varrho(t)-\frac{P_1}{2}\right),
\end{equation}
satisfies eq.~(\ref{equyacobi}) as well, therefore, one and the
same function $y(t)$ corresponds to a two-parameter set of
coefficients $A_i$ ($i=0..4$) and $P_k$ ($k=0..2$), so we always
can put
\begin{equation}
P_2=1 \qquad \mbox{and} \qquad  P_1=0, \label{fix}
\end{equation}
without loss of generality.

In the general case, when one uses Fan's technique~\cite{Fan} (see, for
example,~\cite{Yomba}) with polynomials of degree $L$ in solutions of
eq.~(\ref{equyacobi}) or eq.~(\ref{equwef}), it is possible to simplify
calculations, putting
\begin{equation}
P_L=1 \qquad \mbox{and}  \qquad P_{L-1}=0. \label{fixL}
\end{equation}
In particular, the consideration of linear combinations $P_1y+P_0$
instead of the function~$y$ is useless.

System (\ref{algsyst}) has been solved by REDUCE using the standard function
\textbf{solve} and the Gr\"obner basis method~\cite{VeTish04}. The goal of
this paper is to show that the use of the Laurent series solutions allows to
obtain some solutions of~(\ref{algsyst}) solving only linear systems and
nonlinear equations in one variable.

We cannot use this method for arbitrary $C$, because the Laurent series
solutions are different for different $C$. So, first of all we have to fix
value of $C$. Note that if $A_3=A_1=0$ then $y$ satisfies eq.~(\ref{equwef})
with $\tilde D=0$, hence, such solutions are known. The remaining problem can
be separated on two cases: $A_3\neq 0$ or $A_3=0$ and $A_1 \neq 0$. If
$A_3\neq 0$, then from two first equations of system~(\ref{algsyst}) it
follows that
\begin{equation*}
$$
\displaystyle C=-\:\frac{4}{3}  \quad\quad\mbox{and}\quad\quad
A_4=-\:\frac{4}{3}\qquad \quad\mbox{or} \quad\qquad
C=-\:\frac{16}{5} \quad\quad\mbox{and}\quad\quad
A_4=-\:\frac{32}{15}.
\end{equation*}
Surely, if some solution with $A_3=0$ corresponds to the
above-mentioned values of $C$, then it can be found as well.

\subsection{Construction of linear algebraic system}
Let choose $C={}-4/3$ (the case $C={}-16/5$ one can consider analogously).
After substitution $C={}-4/3$ and $A_4={}-4/3$ in (\ref{algsyst}) we obtain
\begin{equation}
\left\{
\begin{array}{l}
\displaystyle 512P_0+128A_2-256\lambda_1+384\lambda_2+315 A_3^2=0,\\[2.7mm]
\displaystyle 56A_1+(640P_0+195A_2+240\lambda_1+60\lambda_2)A_3=0,\\[2.7mm]
\displaystyle
63A_3A_1+24\left(A_2+4\lambda1+\lambda_2+\frac{32}{3}P_0
\right)A_2+ {} \\[1mm] {} + 256\lambda_1P_0
+640P_0^2+96\lambda_1\lambda_2+288P_0\lambda_2=0,\\[2.7mm]
\displaystyle  10A_3 A_0 + \left(5A_2 - \frac{128}{3}P_0 -
16\lambda_1 - 4\lambda_2
 \right)A_1=0.\\
\end{array} \right.
\label{algsyst43}
\end{equation}

If we consider system~(\ref{algsyst43}) separately from the differential
equations (\ref{equyacobi}) and (\ref{equro}), from which it has been
obtained, then it would be difficult to solve system~(\ref{algsyst43}) without
the use of the Gr\"obner basis method. At the same time from
equations~(\ref{equyacobi}) and (\ref{equro}) we can obtain additional
information, which assists us to solve system~(\ref{algsyst43}).

Let us construct the Laurent series solutions for
 eq.~(\ref{equro}).  The method of construction of the Laurent series
solutions for the generalized H\'enon--Heiles system has been described in
detail in~\cite{VernovTMF}. For eq.~(\ref{equro}) with $C={}-4/3$ we obtain
that solutions have singularities proportional to $1/t$ and the values of
resonances are $-1$ (corresponding to the arbitrary parameter $t_0$), $1$, $4$
and $10$. The Laurent series solutions are (we put $t_0=0$):
\begin{equation}
  \tilde \varrho=\pm\left(\frac{i\sqrt{3}}{t}+c_0+\frac{i\sqrt{3}}{24}
  \left(3\lambda_2-2\lambda_1+4P_0+62c_0^2\right)t+   \dots\right),
\end{equation}
where
\begin{equation}
  c_0=\frac{\pm\sqrt{161700\lambda_1-121275\lambda_2
\pm\sqrt{1155(5481\lambda_2^2-12768\lambda_1\lambda_2+8512\lambda_1^2)}}}{2310}.
\label{C0}
\end{equation}
Two signs $"\pm"$ in (\ref{C0}) are independent. At the same time,
functions $\tilde \varrho$ and $-\tilde\varrho$ correspond to one
and the same function $\tilde y$, so there are four different
Laurent series solutions. The coefficients $c_3$ and $c_9$ are
arbitrary. To find any number of coefficients, we should solve
only two nonlinear equations in one variable and linear equations
in one variable.

The algorithm of the construction of elliptic solutions from the Laurent
series solutions is the following~\cite{CoMu03}:

\begin{itemize}
\item Choose a positive integer $m$ and define the first order ODE
   $(\ref{subequ})$  with unknown coefficients $h_{jk}$.

\item Compute coefficients of the Laurent series $\tilde \varrho$.
The number of coefficients has to be greater than the number of
unknowns.

\item Substituting the obtained coefficients, transform
eq.~$(\ref{subequ})$ into a linear and over\-determined system in
$h_{jk}$ with
   coefficients depending on arbitrary parameters.

\item Exclude $h_{jk}$ and obtain the nonlinear system in
   parameters.

\item Solve the obtained system.
\end{itemize}

To obtain the explicit form of the elliptic function, which satisfies the
known first order ODE, one can use the classical method due to Poincar\'e,
which has been implemented in Maple as the package
``algcurves''~\cite{MapleAlgcurves}.

 On the first step we choose eq.~(\ref{subequ}), which coincides with
eq.~(\ref{equyacobi}). It means that $m=2$, all $h_{j1}$ are equal
to zero and all $h_{j0}={}-A_j/4$. After the second and the third
steps we obtain a linear system in $A_{j}$. The package of
computer algebra procedures, which transforms the first order
equation into a such system of algebraic equations has been
written~\cite{VernovCASC} in Maple.

 The obtained system has the triangular form and is linear in $H$, $c_3$ and
 $c_9$ as well. From the first equation we obtain anew that $A_4={}-4/3$.
From the second equation it follows:
\begin{equation}
A_3=\frac{16}{3}c_0,
\end{equation}
and so on:
\begin{equation}
A_2={}-70c_0^2-3\lambda_2+2\lambda_1-4P_0,
\end{equation}
\begin{equation}
A_1=\left(\frac{40}{3}P_0-60\lambda_1+50\lambda_2+1300c_0^2\right)c_0,
\end{equation}
\begin{equation}
\begin{array}{lcl}
\displaystyle A_0&\displaystyle =&
\displaystyle{}-40i\sqrt{3}c_3-\frac{21535}{12}c_0^4+
\left(\frac{565}{6}\lambda_1 -
\frac{405}{4}\lambda_2-\frac{245}{3}P_0\right)c_0^2+{}\\[2.7mm]
&\displaystyle +&\displaystyle\frac{7}{4}\lambda_1\lambda_2
-\frac{21}{16}\lambda_2^2-\frac{7}{12}\lambda_1^2+\frac{7}{3}{\lambda_1}P_0
-\frac{7}{2}\lambda_2P_0-\frac{7}{3}P_0^2.
\end{array}
\end{equation}
From the next equation of the system we obtain $c_3$ and, finally,
\begin{equation}
\begin{array}{lcl}
\displaystyle A_0&\displaystyle
=&\displaystyle\frac{15645}{4}c_0^{4}+\left(\frac{1545}{4}\lambda_2-465P_0
-{\frac{1495}{2}}\lambda_1\right)c_0^2 +
\frac{537}{20}\lambda_1^2-{}
\\[2.7mm]
&\displaystyle -&\displaystyle\frac{663}{20}\lambda_1\lambda_2+
\frac{729}{80}\lambda_2^2 + 19\lambda_1P_0 -
\frac{37}{2}\lambda_2P_0 -\frac{17}{3}P_0^2.
\end{array}
\end{equation}

Substituting the values of $A_k$, which correspond to one of the
possible values of $c_0$, in system (\ref{algsyst43}) we obtain that
it is satisfied for all values of $\lambda_1$, $\lambda_2$ and
$P_0$, so we do not need to solve the
 nonlinear equations. Therefore we settle the nonlinear algebraic
system~(\ref{algsyst43}), solving only linear equations and nonlinear equation
in one variable. We have used the values of only six coefficients of the
Laurent series solutions. Note that, for $c_0$ and ${}-c_0$ we obtain one and
the same values of $A_4$, $A_2$ and $A_0$, and the opposite values of $A_3$
and $A_1$. From~(\ref{equyacobi}) it follows that these solutions correspond
to $\pm\varrho(t)$, and, hence, give one and the same function $y(t)$.
Therefore not four, but two different elliptic (or degenerated elliptic)
solutions of eq.~(\ref{equy}) have been found. The elliptic solutions $y(t)$
are  fourth-order elliptic functions and can be expressed in terms of the
Weierstrass elliptic function $\wp(t-t_0)$:
\begin{equation}
y(t-t_0) = \left (\frac {a\wp (t-t_0) + b} {c\wp (t-t_0) + d}
\right)^2+P_0, \label{ysol}
\end{equation}
 where constants $a$, $b$, $c$, $d$ and  periods of
$\wp(t)$ are determined by $A_k$. The parameters $t_0$ and $P_0$,
which defines the energy of the system, are arbitrary.
 Solutions of this type exist in both above-mentioned nonintegrable cases:
$C={}-16/5$ and $C={}-4/3$. The full list of solutions is given
in~\cite{VeTish04}.

\section{Multivalued solutions}

 From the Painlev\'e theorem it follows that a solution of an autonomous
polynomial first order differential equation (\ref{1}) can
 have only two types of singular points: poles and algebraic
 branch points. So, the most general type of formal series solutions is the
Puiseux series. Let us generalize the above method by searching for solutions,
which can be expanded in  Puiseux series, as follows:
\begin{equation}
  y=\sum_{k=-L}^{\infty} S_{k/q}t^{k/q}.
\label{puiseux}
\end{equation}

We seek solutions as a polynomial
\begin{equation}
 y=P_L\rho^{L}+P_{L-1}\rho^{L-1}+P_{L-2}
    \rho^{L-2}+P_{L-3}\rho^{L-3}+\dots+P_0.
\end{equation}

From (\ref{puiseux}) it follows that
\begin{equation}
\rho=\sum_{j=-1}^\infty C_{j/q}t^{j/q}.
\end{equation}

We assume that $\rho(t)$ satisfies the following equation:
\begin{equation}
\sum_{k=0}^{m} \sum_{j=0}^{(q+1)(m-k)}h_{jk} \rho^j
{\rho_t}^k=0,\quad h_{0m}=1, \label{puisequ}
\end{equation}
where $q$ is a natural number.

To simplify calculations one can put $P_L=1$ and $P_{L-1}=0$ without loss of
generality, because
\begin{equation}
\breve\rho=\left(\rho-\frac{P_{n-1}}{n}\right)/\sqrt[n]{P_n}
\end{equation}
satisfies (\ref{puisequ}) as well. The construction of the
corresponding  algebraic system has been automatized in
Maple~\cite{VernovPrep04}. We plan to automatize similar
construction for an arbitrary autonomous polynomial first order
equation.

\section{Five--dimensional gravitational model with a scalar field}

 To show how the analysis of singular behavior of solutions can assist to
 find the form of a potential, let us consider a five-dimensional model
 of a gravitational field~\cite{DeWolfe,Gremm}, with the action given by
 \begin{equation}
\tilde S=\int_M d^4x dr \sqrt{|\det\,\tilde
g_{\mu\nu}|}\left(\frac{\tilde R}{4}-\frac{1}{2}\tilde
g^{\mu\nu}\partial_{\mu}\tilde\varphi\partial_{\nu}\tilde\varphi-
V(\tilde\varphi)\right),
\end{equation}
where $M$ is the full five-dimensional space-time. The most
general metric with four-dimensional Poincar\'e symmetry is
\begin{equation}
  ds^2=e^{2A(r)}\left({}-dx_0^2+dx_1^2+dx_2^2+dx_3^2\right)+dr^2.
\end{equation}
Assuming the scalar field depends only on $r$, so that $\phi=\phi(r)$, the
independent equations of motion are
\begin{equation}
\left\{
\begin{array}{lcl}
\displaystyle J'&\displaystyle =&\displaystyle {}-\frac{2}{3}\left(\phi'\right)^2, \\[2.7mm]
\displaystyle  J^2&\displaystyle
=&\displaystyle\frac{1}{6}\left(\phi'\right)^2-\frac{1}{3}V(\phi),
\end{array}
\right.
\label{equh2}
\end{equation}
where $J\equiv A'\equiv \frac{\mathrm{d}A}{\mathrm{d}r}$  and
$\phi'\equiv\frac{\mathrm{d}\phi}{\mathrm{d}r}$.

Similar systems of equations of motion arise also in cosmological models with
a spatially flat Friedmann metric and a scalar field~\cite{Barrow} or a
phantom scalar field~\cite{AKV}.

Let us analyse the correspondence between $\phi(r)$ and the potential
$V(\phi)$. If at a singular point (fixed at the origin, without loss of
generality)
\begin{equation*}
\phi(r)\sim \frac{1}{r^m},
\end{equation*}
then from (\ref{equh2}) it follows that
\begin{equation}
J'\sim\frac{1}{r^{2m+2}} \qquad \Longrightarrow \qquad
J\sim\frac{1}{r^{2m+1}}\qquad \Longrightarrow \qquad V(\phi)\simeq
V\left(\frac{1}{r^m}\right) \sim  \frac{1}{r^{4m+2}}.
\label{27}
\end{equation}

It means that solutions with simple poles can be obtained only if the degree
of the polynomial potential $V(\phi)$ is equal to six. For example, let
$\phi(r)=\tanh(r)$, for real $r$ this function has no singular point, but
system $(\ref{equh2})$ is autonomous, so if $\phi(r)$ is a solution, then
$\phi(r-r_0)$, where $r_0$ is an arbitrary complex constant, is a solution as
well. Then there exists $r_0$ such that $\tanh(r-r_0)$ tends to infinity as
$r\rightarrow 0$, hence, the function $\tanh(r)$ cannot be a solution for any
fourth degree polynomial potential. In~\cite{DeWolfe} the explicit form of the
sixth degree polynomial potential $V(\phi)$, which corresponds to $\tanh(r)$
has been found. Analogously one can show that if solutions tend to infinity as
$1/r^2$, then the degree of $V(\phi)$ is equal to 5. If solutions tend to
infinity as $1/r^k$, where $k$ is a natural number greater than two, then from
(\ref{27}) we obtain that $V(\phi)$ cannot be a polynomial.

The polynomial potentials, which degrees are more than 6, correspond to
multivalued solutions. If solutions tend to infinity as $r^{-2/k}$, then from
(\ref{27}) it is follows that the degree of $V(\phi)$ has to be equal to
$4+k$. For example, if a solution in the neighbourhood of its singular points
tends to infinity as $1/\sqrt{r}$, then $k=4$ and the degree of the
corresponding potential $V(\phi)$ is equal to $8$.

 In conclusion of
this section we say a few words about explicit solutions.
Following~\cite{DeWolfe} we assume that $J(r)$ is a function of~$\phi$:

\begin{equation}
 J(r)=-\:\frac{1}{3}W(\phi(r)).
\end{equation}

 It is straightforward to verify that system (\ref{equh2}) is equivalent to
\begin{equation}
\frac{\mathrm{d}\phi(r)}{\mathrm{d}r}=\frac{1}{2}\,\frac{\mathrm{d}W(\phi)}{\mathrm{d}\phi},
\end{equation}
\begin{equation}
\left(\frac{\mathrm{d}W(\phi)}
{\mathrm{d}\phi}\right)^2-\frac{1}{3}W(\phi)^2-V(\phi)=0.
\label{Wequ}
\end{equation}

Unfortunately the nonautonomous equation~(\ref{Wequ}) can not be solved
analytically. For polynomial $V(\phi)$ it may be possible to find only special
solutions, e.g. $W(\phi)$ in polynomial form. For example, $W(\phi)$ cannot be
a polynomial if $V(\phi)=\left(\phi^2-1\right)^2$ and we don't know a solution
for this potential. At the same time, if the potential
$V(\phi)=A\phi^4+B\phi^2+C$ has no double roots, then eq.~(\ref{Wequ}) has a
polynomial solution.

Contrary to a scalar field theory without gravitational field, there is not a
one-to-one correspondence between the form of the scalar field $\phi(r-r_0)$
and potential $V(\phi)$. The form of the scalar field is defined by
$\frac{\mathrm{d}W}{\mathrm{d}\phi}$, so one can add a constant to $W(\phi)$
and obtain new $V(\phi)$  for the same $\phi(r)$. On the other hand, for given
$V(\phi)$  we have not one-, but two-parameter set of functions~$\phi(r)$.

\section{Conclusion}

The Laurent series solutions are useful to find elliptic or elementary
solutions in the analytic form. The method, proposed in~\cite{CoMu03}, has
been automatized and generalized on some type of multivalued solutions. It
converts the local information into the global one and can be used not only as
an alternative to the standard method, but also as an addition to it, which
assists to find solutions of the obtained algebraic system. We have
demonstrated that one can find elliptic solutions of the generalized
H\'enon--Heiles system  solving only linear equations and nonlinear equations
in one variable, instead of nonlinear system~(\ref{algsyst}). At the same
time, to use this method one has to know not only an algebraic system, but
also the differential equations  from which this system has been obtained.  To
find solutions of the initial ODE in the form of the Laurent or Puiseux series
we use the Painlev\'e test. This test can also assist to solve the inverse
problem: to find the degree of a polynomial potential, which corresponds to
the required type of solutions.

\begin{acknowledgments}
The author is grateful to \ I.Ya.~Aref'eva, \  R.~Conte, \ A.S.~Koshelev, \
M.N.~Smolyakov  and  \ I.P.~Volobuev \ for valuable discussions.  This work
has been supported in part by RFBR grant 05--01--00758,  Russian Fede\-ration
President's grant NSh--1685.2003.2 and by the Scientific Program
``Universities of Russia'' grant 02.02.503.
\end{acknowledgments}

\label{lastpage}
\end{document}